\documentstyle[aps,preprint,floats,tighten]{revtex}
\input{psfig.tex}

\def\vecq{{\bf q}}
\def\vecw{{\bf w}}
\def\veck{{\bf k}}
\def\vecv{{\bf v}}
\def\vectheta{{\bf \vec\theta}}
\def\veckappa{{\bf \vec\kappa}}
\def\hattheta{{\bf \hat \theta}}

\def\hatk{{\bf \hat k}}
\def\hatz{{\bf \hat z}}

\def\VEV#1{{\langle #1 \rangle}}

\long\def\comment#1{}

\begin{document}
\draft
\preprint{\vbox{\hbox{CU-TP-872}
                \hbox{CAL-651}
                \hbox{CfPA-97-th-27}
                \hbox{astro-ph/9801022}
}}
\title{Calculation of the Ostriker-Vishniac Effect in Cold Dark
Matter Models\footnote{This work is dedicated to the memory of
David Schramm.}}

\author{Andrew H. Jaffe\footnote{Email address: jaffe@physics.berkeley.edu}}
\address{Center for Particle Astrophysics, 301 LeConte Hall,
University of California, Berkeley~~94720}
\author{Marc Kamionkowski\footnote{Email address: kamion@phys.columbia.edu}}
\address{Department of Physics, Columbia University, 538 West
120th Street, New York, New York~~10027}
\date{December 1997}
\maketitle

\begin{abstract}
We present a new derivation of the cosmic microwave background
anisotropy spectrum from the Ostriker-Vishniac effect for an
open, flat, or closed Universe, and calculate the anisotropy
expected in cold dark-matter (CDM) models.  We provide simple
semi-analytic fitting formulas for the Vishniac power
spectrum that can be used to evaluate the expected anisotropy in
CDM models for any arbitrary ionization history. In a flat Universe, CDM
models  normalized to
cluster abundances produce rms temperature anisotropies of 0.8--2.4 $\mu$K
on arcminute angular scales for a constant
ionization fraction of unity, whereas an ionization fraction of
0.2 yields rms anisotropies of 0.3--0.8 $\mu$K. In an open
and/or high-baryon-density Universe, the level of anisotropy is
somewhat higher.  The signal in some
of these models may be detectable with planned interferometry
experiments.  The damping of the acoustic peaks in the
primary-anisotropy spectrum at degree angular scales depends
primarily on the optical depth and only secondarily on the epoch
of reionization.  On the other hand, the amplitude of
Ostriker-Vishniac anisotropies depends sensitively on the epoch
of reionization.  Therefore, when combined with the estimate of the
reionization optical depth provided by maps of degree-scale
anisotropies, the Ostriker-Vishniac effect can provide a unique
probe of the epoch of reionization.
\end{abstract}


\section{INTRODUCTION}

Standard cold dark matter (CDM) and its variants are
the current leading models for the origin of large-scale structure.
The canonical CDM model hypothesizes a primordial scale-free
spectrum of primordial adiabatic perturbations and a critical
density of cold dark matter.  Its variants include low-density
(e.g., $\Omega_0\simeq0.3-0.4$) models, either in an open Universe
or in a flat Universe with a cosmological constant, tilted
models in which the power-law index $n$ of the primordial power
spectrum differs slightly from scale-free ($n=1$),
low--Hubble-constant models, or mixed-dark-matter models in
which the Universe has a critical density but roughly 30\% is in
the form of hot dark matter.

Although most of the matter in these models does not undergo
gravitational collapse until relatively late in the history of
the Universe, some small fraction of the mass is expected to
collapse at early times.  Ultraviolet photons released by this
early generation of star and/or galaxy formation will partially
reionize the Universe, and these ionized electrons
will re-scatter at least some cosmic microwave background (CMB)
photons after recombination
at a redshift of $z\simeq1100$.  Theoretical
uncertainties in the process of star formation and the resulting
ionization make precise predictions of the ionization history
difficult.  Constraints to the shape of the CMB blackbody
spectrum and detection of CMB anisotropy at degree angular
scales suggest that if reionization occurred, the fraction of
CMB photons that re-scattered is small.  Still, estimates show
that even if small, at least some reionization is expected in
CDM models \cite{kamspergelsug,blanchard,haiman}: for
example, the most careful recent calculations suggest a
fraction $\tau_r\sim0.1$ of CMB photons were re-scattered
\cite{haiman}.

Scattering of CMB photons from ionized
clouds will lead to anisotropies at arcminute separations below
the Silk-damping scale (the Ostriker-Vishniac effect)
\cite{ostriker,vishniac,efstathiou,dodelson,hss,persi,huwhite}.  These
anisotropies arise at a higher order in perturbation theory and
are therefore not included in the usual Boltzmann calculations
of CMB anisotropy spectra.  The level of anisotropy is
expected to be small and it has so far eluded detection.  However,
these anisotropies may be observable with forthcoming CMB
interferometry experiments \cite{interferometers} that probe the
CMB power spectrum at arcminute scales.

In this paper, we present a new derivation of
the anisotropy spectrum from the Ostriker-Vishniac effect.  We
then calculate the anisotropy expected in CDM models.  We
provide numerical results for the anisotropy
predicted by a variety of CDM models with several plausible
ionization histories.  We also provide simple semi-analytic
fitting formulas for the Ostriker-Vishniac power spectrum that
can be used to evaluate the expected anisotropy in CDM models
for any arbitrary ionization history.  We estimate the
anisotropy amplitude that may be detectable with forthcoming
interferometry experiments.  We find that the signal should be
detectable in a number of CDM models with plausible ionization
histories.

To a good approximation, the damping of the acoustic peaks in
the primary (linear-theory) CMB spectrum from reionization is
the same for any
ionization history that produces the same optical depth to the
surface of last scatter (see, e.g., Ref. \cite{kamspergelsug}).
Therefore, although MAP and Planck \cite{MAP} should provide a good
estimate of this optical depth \cite{jkks}, they will
not strongly constrain the
ionization history, that is, whether reionization occurred
earlier or later.  On the other hand, our results indicate that
the amplitude of the Ostriker-Vishniac anisotropy depends
primarily on the ionization fraction and only secondarily on the
optical depth.  Therefore, when combined with the determination
of the optical depth from MAP and Planck, the Ostriker-Vishniac
effect can determine the epoch of reionization.  This will
be important for precise determination of the baryon density,
Hubble constant, and cosmological constant from the higher
acoustic peaks in the CMB spectrum.  It will also provide a
unique window to the epoch at which structures first undergo
gravitational collapse in the Universe.

\section{PRELIMINARIES}

The  scale factor of the Universe, $a(t)$, satisfies the
Friedmann equations,
\begin{equation}
     {\dot a \over a} = H_0 E(z) \equiv H_0 \sqrt{\Omega_0 (1+z)^3 +
     \Omega_\Lambda + (1-\Omega_0-\Omega_\Lambda)(1+z)^2},
\end{equation}
\begin{equation}
     {\ddot a \over a} = H_0^2 [\Omega_\Lambda - \Omega_0
     (1+z)^3/2],
\end{equation}
where $H_0 = 100\, h$ km~sec$^{-1}$~Mpc$^{-1}$ is the Hubble
constant, $\Omega_0$ is the current nonrelativistic-matter
density in units of the critical density, $\Omega_\Lambda$ is
the current contribution of the cosmological constant to
the critical density, and the overdot denotes a derivative with respect to
time.  If $\Omega_0+\Omega_\Lambda=1$, the Universe is flat; if
it is greater (less) than unity the Universe is closed (open). An
Einstein-de Sitter Universe has $\Omega_0=1$ and $\Omega_\Lambda=0$, so
$E(z)=(1+z)^{3/2}$.

We choose the scale factor such that $a_0 H_0=2$.  If we are
located at the origin, $\vecw=0$, then an object at redshift $z$
is at a comoving distance,
\begin{equation}
     w(z) = {1\over 2} \int_0^z \, {dz' \over E(z')}.
\end{equation}
We define a conformal time by $d\eta=dt/a$, so the comoving
distance to the horizon is the conformal time today,
$\eta_0=w(\infty)$. In an Einstein-de Sitter Universe,
$a/a_0=\eta^2$ and $\eta_0=1$ with our conventions.

If the density contrast at comoving position $\vecw$ at time $t$
is $\delta(\vecw,t)$, with Fourier transform
\begin{equation}
     \tilde \delta(\veck) = \int \,d^3 \vecw\,\exp(-i \veck \cdot \vecw)\,
     \delta(\vecw)  \qquad \delta(\vecw) = \int \,{d^3 \veck
     \over (2\pi)^3} \,\exp(i
     \veck \cdot \vecw)\, \tilde \delta(\veck),
\end{equation}
then the power spectrum $P(k,t)$ is defined by the expectation value
\begin{equation}
     \VEV{\tilde \delta(\veck,t) \tilde \delta^*(\veck',t)} = (2\pi)^3
     \delta(\veck-\veck') P(k,t),
\end{equation}
where the angle brackets denote an average over all realizations.
In linear theory, the spatial and time dependence of $\delta$
can be factorized, so $\delta(\vecw,t)=\delta_0(\vecw)D(t)/D(t_0)$,
where $t_0$ is the age of the Universe,
$\delta_0(\vecw)=\delta(\vecw,t_0)$, and the growth factor
(written as a function of redshift) is \cite{peebles}
\begin{equation}
     D(z) = {5 \Omega_0\, E(z) \over 2} \, \int_z^\infty \, {1+z' \over
     [E(z')]^3} \, dz'.
\end{equation}
The time $t$ is related to the redshift by
\begin{equation}
     t(z) = {1 \over H_0} \int_z^\infty \, {dz' \over (1+z')
     E(z')}.
\end{equation}
Therefore, the power spectrum is $P(k,t)=P(k,0)(D/D_0)^2$, where
$D_0\equiv D(t_0)$.  In an Einstein-de Sitter Universe,
$a/a_0=(t/t_0)^{2/3}$, and $D(t)/D_0 = a(t)/a_0$.

In linear theory, the Fourier components of the velocity field
are related to those of the density field by
\begin{equation}
     \tilde\vecv(\veck,t) = {i a(t) \over k^2}\, {\dot D \over D}\, \veck
     \, \tilde\delta(\veck, t) = {i a(t) \over k^2}\, {\dot D
     \over D_0} \,\veck \, \tilde \delta_0(\veck).
\label{eq:veckeqn}
\end{equation}
It can also be shown \cite{peebles} that
\begin{equation}
     {\dot D \over D} = {\ddot a \over \dot a} - {\dot a
     \over a} + {5 \Omega_0 \over 2} {\dot a \over a} {(1+z)^2
     \over [E(z)]^2\,D(z)}.
\end{equation}

\section{THE OSTRIKER-VISHNIAC EFFECT}

The Ostriker-Vishniac effect is the CMB anisotropy produced by
scattering from ionized regions or clouds with bulk peculiar velocities.
The fractional temperature perturbation in the direction $\hattheta$
induced by bulk motions is
\begin{equation}
     p(\vectheta) \equiv {\Delta T \over T}(\vectheta) =
     -\int_0^{\eta_0} \, n_e \sigma_T e^{-\tau}
     [\hattheta \cdot \vecv(w\hattheta;w)/c] a(w) dw,
\label{eq:startingpoint}
\end{equation}
where $n_e$ is the electron density along the line of sight,
$\vecv(\vecw;w)$ is the bulk velocity at position $\vecw$ at a
conformal time $\eta_0-w$, $\sigma_T$ is the Thomson cross section, and
$\tau$ is the optical depth from us to $w$.  The factor of
$a(x)$ arises because the physical time is $dt = a(x) dx$.  [Note that
$\hattheta$ represents a three-dimensional unit vector along the line
of sight, whereas $\vectheta$ will refer to a dimensionless
two-dimensional vector in the plane perpendicular to the line of
sight---i.e., for directions near $\hatz$, $\hattheta =
(\theta_1, \theta_2, \sqrt{1-\theta_1^2-\theta_2^2}) \simeq
(\theta_1,\theta_2,1)$, whereas $\vectheta=(\theta_1,\theta_2,0)$.]
The visibility function,
\begin{equation}
     g(w) = \overline{n_e}(w) \sigma_T a(w) e^{-\tau(w)} =
     (d\tau/dw) e^{-\tau}
\end{equation}
is the probability distribution for scattering from reionized electrons.
Here, $\overline{n_e}(w)=\Omega_b \rho_c x_e(w) (1+z)^3/m_p$ is the mean
electron density, where $\Omega_b$ is the baryon density today in units
of the critical density (actually, this number should be multiplied by 7/8
to account for the neutrons bound in helium nuclei); $\rho_c$ is the
critical density; $x_e(w)$ is the ionization fraction; and $m_p$ is the
proton mass.  The visibility function is normalized so that
\begin{equation}
     \int_0^{\infty}\, g(w) dw = 1-e^{-\tau_r},
\label{eq:normalization}
\end{equation}
where $\tau_r$ is the optical depth to the
standard-recombination surface of last scatter, the fraction of
CMB photons which scattered after standard recombination at
$z\simeq1100$.

With these developments, the fractional temperature perturbation
becomes
\begin{equation}
     p(\vectheta) = - \int_0^{\eta_0} \, dw \,
     g(w)\, \hattheta \cdot \vecq(w\hattheta,w),
\label{eq:intermediateone}
\end{equation}
where $\vecq(\vecw,w) = [1+\delta(\vecw,w)] \vecv(\vecw,w)$.
According to Eq.~(\ref{eq:intermediateone}), only the component
of $\vecv$ along the line of sight contributes to the
anisotropy.  In linear theory, we neglect $\delta\ll 1$ and the
total anisotropy is the sum of the contributions of each Fourier
component of the velocity field.  However, in linear theory (and
in fact to all orders), $\tilde\vecv(\veck) \propto \veck$, so only
$\veck$ modes along the line of sight can contribute to the
anisotropy.  The contributions of troughs and crests of
each Fourier mode cancel approximately when projected along the
line of sight.  Therefore, the contribution of bulk velocities
to anisotropies vanishes (or is very small) to lowest order
in perturbation theory on small angular scales.  On larger
angular scales, there are fewer crests and troughs, so the
cancellation is not as complete.

In other words, angular correlations (on small
angular scales) can be due only to $\veck$ modes perpendicular to
the line of sight, and these contribute nothing to the
anisotropy from bulk velocities alone.  Furthermore, no
anisotropy can be produced by $\vecq$ modes parallel to $\veck$,
and therefore the anisotropy can be due only to the components
$\tilde\vecq_\perp(\veck)$ perpendicular to $\veck$
\cite{ostriker,vishniac,dodelson,hss}.

In the Appendix, we make this argument more precise and show that the
projection, Eq.~(\ref{eq:intermediateone}), has Fourier coefficients
that satisfy
\begin{equation}
     \VEV{ \tilde p(\vec\kappa)  \tilde p^*(\vec\kappa')} = (2 \pi)^2
     \delta(\vec\kappa - \vec\kappa') P_p(\kappa),
\end{equation}
with
\begin{equation}
     P_p(\kappa) = {1\over 2} \int_0^{\eta_0} \, {g^2(w) \over w^2}\,
     P_\perp( \kappa/w,w)\, dw,
\label{eq:Ppequation}
\end{equation}
in a flat Universe [making the replacement
Eq.~(\ref{eq:replacement}) for a closed or open Universe] where
$P_\perp(k,w)$ is the three-dimensional power spectrum for
$\tilde\vecq_\perp(\veck)$.  We now need to determine this
three-dimensional power spectrum.

Again, $\tilde\vecv(\veck) \propto \veck$, so $\tilde\vecq_\perp(\veck)$ can
only come from the Fourier transform of the nonlinear term
$\delta(\vecw) \vecv(\vecw)$.  The Fourier transform of
$\delta(\vecw,t) \vecv(\vecw,t)$ [not the full $(1+\delta)\vecv$] is
\begin{eqnarray}
     \tilde\vecq(\veck,t) & = & {1\over2}\int\ {d^3 \veck' \over (2 \pi)^3}\,
     \left[ \tilde \delta(\veck',t)\tilde \vecv(\veck-\veck',t)
     +\tilde \delta(\veck-\veck',t)\tilde\vecv(\veck',t) \right] \nonumber \\
     & =& {i a \dot D D \over 2 D_0^2} \int\ {d^3 \veck' \over (2
     \pi)^3}\, \tilde \delta_0(\veck') \tilde\delta_0(\veck-\veck') \left(
     {\veck -\veck' \over |\veck -\veck'|^2} + { \veck' \over
     k'^2} \right).
\end{eqnarray}
The components of $\tilde\vecq$ perpendicular to $\veck$ are
\begin{eqnarray}
     q_{\perp,i}(\veck,t) & = & \left(\delta_{ij} - {k_i k_j \over
     k^2} \right) q_j(\veck,t) \nonumber \\
     & = & {i a \dot D D \over 2 D_0^2} \int\ {d^3 \veck' \over (2
     \pi)^3}\, \tilde\delta_0(\veck') \tilde\delta_0(\veck-\veck')
     \left(k'_i - {k_i (\veck\cdot \veck') \over k^2}
     \right) \left( { 1 \over k'^2} - {1 \over |\veck
     -\veck'|^2} \right).
\end{eqnarray}
For a Gaussian density field,
\begin{eqnarray}
     &&\VEV{\tilde\delta_0(\veck_1 - \veck') \tilde \delta_0(\veck')
     \tilde \delta_0^*( \veck_2-\veck'')
     \tilde\delta_0^*(\veck'')} \nonumber \\
     &&= (2 \pi)^6 P(|\veck_1 - \veck'|,t) P(k',t) [ \delta( \veck_1 -
     \veck_2) \delta(\veck'-\veck'') + \delta(\veck_1 - \veck_2)
     \delta(\veck_1 - \veck' -\veck'')].
\end{eqnarray}
We combine these expressions to find (after some extensive although
straightforward algebra)
\begin{equation}
     \VEV{\tilde\vecq_\perp(\veck,t) \cdot\tilde\vecq_\perp^*(\veck',t)} = (2
     \pi)^3 \, \delta(\veck-\veck') P_\perp(\veck,t),
\end{equation}
with
\begin{equation}
     P_\perp(k,w) = {a^2(w) \over 8 \pi^2} \left( {\dot D D \over
     D_0^2} \right)^2 S(k);
\label{eq:Pperp}
\end{equation}
the Vishniac power spectrum is \cite{vishniac}
\begin{equation}
     S(k) = k \int_0^\infty dy \int_{-1}^1 dx P(ky) P(k\sqrt{1+y^2
     -2xy}) { (1-x^2)(1-2xy)^2 \over (1+y^2 -2xy)^2}.
\label{eq:Sofk}
\end{equation}
(The differences with Eq.~(2.13) in \cite{vishniac} are the
result of differing Fourier conventions.)

Assembling these results, in a flat Universe, the
Ostriker-Vishniac effect produces a power spectrum,
\begin{equation}
     P_p(\kappa) = {1 \over 16 \pi^2}\, \int_0^{\eta_0}\, {g^2(w)
     \over w^2} \, [a(w)]^2 \left( {\dot D D \over
     D_0^2} \right)^2 S(\kappa/w)\, dw,
\label{eq:vishniacresult}
\end{equation}
with the Vishniac power spectrum $S(k)$ given by
Eq.~(\ref{eq:Sofk}).  
In an open or closed Universe, one makes the replacement
Eq.~(\ref{eq:replacement}) in the argument of $S(\kappa/w)$ and in the
two factors of $w$ in the denominator of the integrand.the numbers for
the open models in the table and the open-model curves need to be fixed.

\subsection{Gaussian Visibility Function}

If the visibility function is approximated as a Gaussian in
conformal time,
\begin{equation}
     g(w) = {1-e^{-\tau_r} \over \sqrt{2 \pi (\Delta w)^2}} \exp \left[
     -{1\over2} { (w-w_r)^2 \over (\Delta w)^2} \right],
\end{equation}
where $w_r$ is the comoving distance to the re-scattering
surface, $\Delta w$ is the width of the re-scattering surface,
and $\tau_r$ is the fraction of CMB photons re-scattered, then
the CMB power spectrum is
\begin{equation}
     P_p(\kappa) \simeq { \left(1-e^{-\tau_r}\right)^2 \over 32
     \pi^2 \sqrt{\pi} (\Delta
     w) w_r^2} \left[ {a(w_r) \dot D(w_r) D(w_r) \over D_0^2}
     \right]^2 S(\kappa/w_r).
\end{equation}
In practice, the visibility functions for the ionization
histories usually considered are very poorly approximated by a
Gaussian.  Still, the Gaussian approximation provides a simple
analytic description of how the anisotropy spectrum depends on
the redshift of re-scattering, optical depth, and $S(k)$.

\subsection{Relation to Multipole Moments}

Given the power spectrum $P_p(\kappa)$ and our Fourier
conventions, it is straightforward to write the temperature
autocorrelation function in terms of the power spectrum:
\begin{equation}
     C(\alpha) = \VEV{p(\vectheta+\vec\alpha) p(\vectheta)} =
     {1\over 2 \pi} \, \int\, \kappa \, d\kappa\, J_0(\kappa
     \alpha)\, P_p(\kappa).
\end{equation}
The correlation function can also be written in terms of the
more commonly used CMB multipole moments $C_\ell$ as
\begin{equation}
     C(\alpha) = \sum_\ell {2\ell+1 \over 4 \pi} \, P_\ell(\cos\alpha) C_\ell,
\end{equation}
where $P_\ell(\cos\alpha)$ are Legendre polynomials.  Since
$P_\ell(\cos\theta) \rightarrow J_0(\ell\theta)$ as $\ell\rightarrow
\infty$, it follows that
\begin{equation}
     C_\ell = P_p(\kappa=\ell).
\label{eq:ClPprelation}
\end{equation}
This identification allows us to illustrate our results in the
multipole moments that have become familiar and to compare our
results with those of previous authors.

\subsection{Comparison with Previous Results}

First, we compare with the results of Ref.\cite{dodelson} (and
therefore Ref.\cite{efstathiou}).  To do so, we consider only the
Einstein-de Sitter Universe as they do.  In an Einstein-de
Sitter Universe, Eq.~(\ref{eq:vishniacresult}) becomes (after
changing the integration variable to $\eta=\eta_0-w$),
\begin{equation}
     P_p(\kappa) = {1 \over 4 \pi^2} \int_0^{\eta_0} \,
     {g^2(w=\eta_0-\eta) \over (\eta_0-\eta)^2 \eta_0^2} \left(
     {\eta \over \eta_0} \right)^6 S\left( {\kappa \over \eta_0
     -\eta} \right)\, d\eta.
\label{eq:EdSresult}
\end{equation}
We then note that they approximate the power spectrum by
evaluating $S(\kappa/w)$ at $w=\eta_0$ and taking it outside the
integral in Eq.~(\ref{eq:EdSresult}).  Now we compare our result with
Eq.~(6.15) in Ref.\cite{dodelson} and their $J$ in their Eq.~(6.18).  If
we note that their $K(k)P^2(k)$ is equal to our $S(k)/k$ and
assume that $\eta_0-\eta \simeq \eta_0$ in the integrand, then
our results reproduce theirs (although it appears that
their result is twice ours; we have not been able to locate the
source  of the discrepancy).

Now we want to compare our results with \cite{huwhite}.
Although their derivation,
unlike ours, relies on a solution to the Boltzmann equation,
our results should agree.  Specifically, our
Eq.~(\ref{eq:vishniacresult}) should agree with their Eq.~(24).  To
check that this is so, we change our integration variable in
Eq.~(\ref{eq:vishniacresult}) to $k=\ell/w$ and find
\begin{equation}
     C_\ell = \int_{\eta_0}^\infty\, dk \, W_\ell(k) \, S(k),
\end{equation}
where
\begin{equation}
     W_\ell(k) = {1\over 16 \pi^2 \ell} \left[g(w=\ell/k)\right]^2 \left(
       {a \dot D D \over D_0^2} \right)^2.
\end{equation}
If we put their Eq.~(24) into this form [to do this, keep in mind
that their overdot denotes differentiation with respect to $\eta$
instead of $t$, so our $a \dot D$ is equal to their $\dot
D$---also, they normalize $D$ such that $D(t_0)=1$], we find
agreement with their results.

\subsection{Variance of Temperature Distribution}

The variance of the temperature distribution induced by the
Ostriker-Vishniac effect is
\begin{equation}
     \VEV{(\Delta T/T)^2} = {1 \over 2 \pi} \int\, \ell \, d\ell\, C_\ell.
\end{equation}
In the Gaussian approximation for $g(w)$ and in a flat Universe,
\begin{equation}
     \VEV{(\Delta T/T)^2} = {\tau_r^2 \over 32 \pi^3 \sqrt{\pi} (\Delta w)}
     \left[ { a(w_r) \dot D(w_r) D(w_r) \over D_0^2} \right]^2
     \, \int \, k \, dk\, S(k),
\label{eq:gaussianvariance}
\end{equation}
and for an Einstein-de Sitter Universe this simplifies to
\begin{equation}
     \VEV{(\Delta T/T)^2} = {\tau_r^2 \over 32 \pi^3 \sqrt{\pi} (\Delta w)}
     { 4 \over \eta_0^2} \left({\eta_r \over \eta_0}\right)^6 \, \int
     \, k\, dk\, S(k).
\end{equation}

\section{THE SPATIAL POWER SPECTRUM}

For the power spectrum, we use
\begin{eqnarray}
     P(k) &=& {2 \pi^2 \over 8} \delta_H^2 (k/2)^n T^2(k_p\,{\rm
     Mpc}/h \Gamma),
\label{eq:powerspectrum}
\end{eqnarray}
where $T(q)$ is the usual CDM transfer function, $k_p= k a_0 =k
H_0/2$ is the physical wave number with our conventions, and
$\Gamma \simeq \Omega_0
h$ is given more accurately in terms of $\Omega_0 h$ and the
baryon fraction $\Omega_b$ by Eqs.~(D-28) and (E-12) in
Ref.\cite{husugiyama}.
The factor of 8 in the denominator in
Eq.~(\ref{eq:powerspectrum}) arises because we are using $a_0
H_0=2$.  For the transfer function, we use \cite{bar86}
\begin{equation}
     T(q)= {\ln(1+2.34q)/(2.34 q)\over [1+3.89 q + (16.1 q)^2 +
     (5.46 q)^3 + (6.71 q)^4]^{1/4}}.
\label{eq:transferfunction}
\end{equation}
For $\delta_H$, we use the fits to the {\sl COBE} anisotropy
given in \cite{bunn}.  For the tilted model, we use the
{\sl COBE} normalization from Ref. \cite{bunn} obtained
including tensor modes with the amplitude predicted by power-law
inflation.  (Note that the tensor contribution in other
inflationary models may be different.)

Alternatively, the power
spectrum may be normalized at small distance scales through the
cluster abundance which fixes $\sigma_8$, the variance in the
mass enclosed in spheres of radius $8\,h^{-1}$~Mpc, to
$\sigma_8\simeq (0.6 \pm 0.1)\Omega_0^{-0.53}$ \cite{via96}.  In
terms of the power spectrum,
\begin{equation}
     \sigma_8^2 = {1 \over 2 \pi^2} \int\,k^2 \, dk \, P(k)
     \left[ { 3 j_1(k_p R) \over k_p R} \right]^2,
\end{equation}
where $R=8\,h^{-1}$ Mpc, and $j_1(x)$ is a spherical Bessel
function.  Since we are using $a_0\neq1$, $k_p$ (rather than
$k$) enters into the argument of the spherical Bessel function.

\section{NUMERICAL RESULTS}

In this Section we present numerical results for the anisotropy
spectrum and rms temperature fluctuation for a variety
of CDM models and ionization histories.  We consider a family of
ionization histories parameterized by a constant ionization
fraction $x_e$ after a redshift $z_r$.  In a cosmological-constant
Universe, this induces an optical depth,
\begin{eqnarray}
     \tau_r &=& (\Omega_b \rho_c \sigma_T a_0/m_p) \,
     \int_0^{z_r} \, { (1+z)^2 \, dz \over E(z)} \nonumber \\
     &=& 0.069\,\Omega_b h x_e \int_0^{z_r} \, { (1+z)^2 \,
     dz \over E(z)}.
\label{eq:tauequation}
\end{eqnarray}
In a flat Universe, this becomes
\begin{equation}
     \tau_r = 0.046\, \Omega_b h x_e \, \left[
     \sqrt{\Omega_0(1+z_r)^3 + (1-\Omega_0)} -1 \right].
\label{eq:flattauequation}
\end{equation}
The prefactor 0.069 (or 0.046 in the flat-Universe expression)
is obtained if all the baryons are in the form of protons,
an assumption often made when calculating ionization histories.
If one takes into account the fact that one quarter of the
baryonic mass is helium, then the prefactor should be multiplied
by 7/8.

\begin{table*}
\caption{The rms temperature perturbation $\VEV{ (\Delta
  T/T)^2}^{1/2}$ in $\mu$K due to the Ostriker-Vishniac effect and value
  of $\ell_{\rm peak}$ at which $\ell(\ell+1) C_\ell$ peaks for a variety of
  CDM models and ionization histories for models normalized to both {\sl
    COBE} and cluster abundances.}
\begin{center}
\begin{tabular}{|c|c|c|c|c|c|c|c|c|c|c|c|c|}
 $\Omega_0$ & $\Omega_\Lambda$& $h$ & $n$ & $\Omega_b h^2$ &
 $\sigma_8$ & $\sigma_8 \Omega_0^{0.53}$ & $z_r$ & $x_e$
 & $\tau_r$ & $(\Delta T)_{\rm rms}^{\rm COBE}$ & $(\Delta
 T)_{\rm rms}^{\rm cluster}$ & $\ell_{\rm peak}$  \\
 \hline \hline
 1 & 0 & 0.5 & 1 & 0.0125 & 1.21 & 1.21 & 5   & 1.0 & 0.016 & 3.1& 0.77 & 4000 \\ \hline
 1 & 0 & 0.5 & 1 & 0.0125 & 1.21 & 1.21 & 19  & 1.0   & 0.1 & 4.8& 1.2  & 6300 \\ \hline
 1 & 0 & 0.5 & 1 & 0.0125 & 1.21 & 1.21 & 56  & 0.2   & 0.1 & 1.4& 0.34 & 7900\\ \hline
 1 & 0 & 0.5 & 1 & 0.0125 & 1.21 & 1.21 & 56  & 1.0   & 0.5 & 6.2& 1.5  & 7900 \\ \hline
 1 & 0 & 0.5 & 1 & 0.0125 & 1.21 & 1.21 & 166 & 0.2   & 0.5 & 1.7&  0.42& 7900 \\ \hline

 1 & 0 & 0.5 & 0.8 & 0.025 & 0.53 & 0.53 & 12  & 1.0  & 0.1 & 1.2  & 1.6 & 3200\\\hline
 1 & 0 & 0.5 & 0.8 & 0.025 & 0.53 & 0.53 & 35  & 0.2  & 0.1 & 0.37 & 0.48& 4000 \\ \hline
 1 & 0 & 0.5 & 0.8 & 0.025 & 0.53 & 0.53 & 35  & 1.0  & 0.5 & 1.6  & 2.1 & 4000 \\ \hline
 1 & 0 & 0.5 & 0.8 & 0.025 & 0.53 & 0.53 & 104 & 0.2  & 0.5 & 0.46 & 0.60& 4000\\ \hline

 0.4 & 0.6 & 0.65 & 1 & 0.015 & 1.07 & 0.65 & 27  & 1.0 & 0.1 &   2.5 & 2.1 & 4000\\ \hline
 0.4 & 0.6 & 0.65 & 1 & 0.015 & 1.07 & 0.65 & 81  & 0.2 & 0.1 &   0.73& 0.62& 5000 \\ \hline
 0.4 & 0.6 & 0.65 & 1 & 0.015 & 1.07 & 0.65 & 81  & 1.0 & 0.5 &   3.3 & 2.8 & 5000 \\ \hline
 0.4 & 0.6 & 0.65 & 1 & 0.015 & 1.07 & 0.65 & 238 & 0.2 & 0.5 &   0.92& 0.77& 6300 \\ \hline

 1 & 0 & 0.35 & 1 & 0.015 & 0.74 & 0.74 & 13   & 1.0  & 0.1  & 2.4 &   1.6 & 4000\\ \hline
 1 & 0 & 0.35 & 1 & 0.015 & 0.74 & 0.74 & 39   & 0.2  & 0.1  & 0.70&   0.46& 5000\\ \hline
 1 & 0 & 0.35 & 1 & 0.015 & 0.74 & 0.74 & 39   & 1.0  & 0.5  & 3.1 &   2.1 & 4000 \\ \hline
 1 & 0 & 0.35 & 1 & 0.015 & 0.74 & 0.74 & 116  & 0.2  & 0.5  & 0.87&   0.57& 5000 \\ \hline

 0.4 & 0 & 0.8 & 1 & 0.0125 & 0.87& 0.54 & 19  & 1.0  & 0.1 & 2.3 &   2.8 & 7900\\ \hline
 0.4 & 0 & 0.8 & 1 & 0.0125 & 0.87& 0.54 & 54  & 0.2  & 0.1 & 0.79&  0.99 &10000\\ \hline
 0.4 & 0 & 0.8 & 1 & 0.0125 & 0.87& 0.54 & 54  & 1.0  & 0.5 & 3.2 &   4.0 &10000\\ \hline
 0.4 & 0 & 0.8 & 1 & 0.0125 & 0.87& 0.54 & 156 & 0.2  & 0.5 & 1.0 &   1.3 &10000\\ \hline

 0.3 & 0 & 0.7 & 1 & 0.05 & 0.30& 0.16 & 7    & 1.0  & 0.1 & 0.79 &   11  & 2500\\ \hline
 0.3 & 0 & 0.7 & 1 & 0.05 & 0.30& 0.16 & 20   & 0.2  & 0.1 & 0.31 &   4.4 & 4000\\ \hline
 0.3 & 0 & 0.7 & 1 & 0.05 & 0.30& 0.16 & 20   & 1.0  & 0.5 & 1.3  &   19  & 3100\\ \hline
 0.3 & 0 & 0.7 & 1 & 0.05 & 0.30& 0.16 & 57   & 0.2  & 0.5 & 0.44 &   6.4 & 5000\\ 
\end{tabular}
\end{center}
\label{tab:predictions}
\end{table*}

Table~\ref{tab:predictions} lists the predicted rms temperature
anisotropy $\VEV{ (\Delta T)^2}^{1/2}$ due to the
Ostriker-Vishniac effect for a variety of CDM models,
which are parameterized by $\Omega_0$, $\Omega_\Lambda$, $h$,
$n$, and $\Omega_b h^2$, and
ionization histories, which are parameterized by $x_e$ and $z_r$.  The
Table also lists $\tau_r$, $\sigma_8$, and $\sigma_8
\Omega_0^{0.53}$.  We present a variety of models and ionization
histories to illustrate the effects of variation of various
parameters on the predictions.   It should be kept in mind that
not all of the {\sl COBE}-normalized models satisfy the
aforementioned observational constraint to $\sigma_8$ from
cluster abundances \cite{via96}.  We therefore list in the Table
the predicted RMS temperature anisotropy expected if the {\sl COBE}
normalization is disregarded and the amplitudes of the power
spectrum is fixed instead to the cluster abundance.
Similarly, the $h=0.35$ model is meant to be illustrative; this
value of the Hubble constant is in disagreement with most
measurements.

\begin{figure}[htbp]
\centerline{\psfig{file=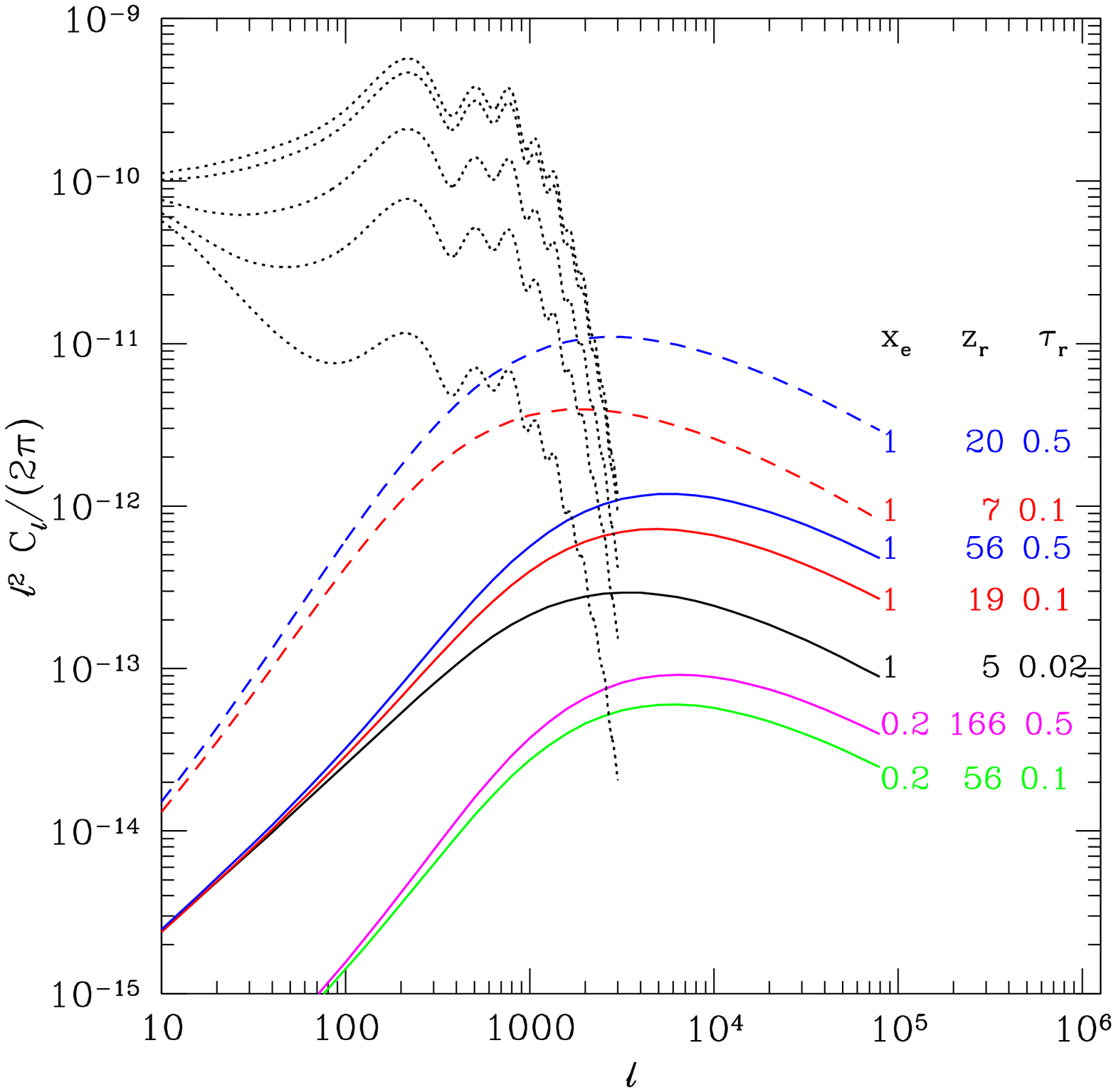,width=5in}}
\bigskip
\caption{
  Multipole moments for the Ostriker-Vishniac effect for the
  {\sl COBE}-normalized canonical standard-CDM model ($\Omega=1$, $h=0.5$,
  $n=1$, $\Omega_b h^2 =0.0125$), for a variety of ionization
     histories, as listed.
     We also show predictions for several open high-baryon-density models with
     the same $x_e$ and $\tau_r$,
     normalized to the cluster abundance, with dashed curves.  The dotted
     curves show the primary anisotropy for this model for $\tau_r=0.0$,
     0.1, 0.5, 1, and 2, from top to bottom.  }
\label{fig:Clsplot}
\end{figure}

Figure~\ref{fig:Clsplot} shows the predicted anisotropy for the
canonical standard-CDM model ($\Omega_0=1$, $\Omega_\Lambda=0$, $h=0.5$,
$n=1$, $\Omega_b h^2 =0.0125$), for a variety of ionization histories,
with the linear power spectrum, also shown, normalized to {\sl COBE}.
We know from quasar absorption spectra that the Universe has been
significantly reionized at least since a redshift of $z_r=5$.
Therefore, we include predictions for this minimal level of anisotropy.

The effect of reionization on the primary anisotropies can be quantified
primarily by the optical depth $\tau_r$\cite{kamspergelsug}; any two
ionization histories that have the same $\tau_r$ produce roughly the
same primary anisotropies.  However, as Fig.~\ref{fig:Clsplot} and
Table~\ref{tab:predictions} show, the secondary anisotropies are more
dependent on the ionization fraction than on the optical depth.

The growth of density perturbations leads to a
growth of the peculiar velocities that induce the secondary
anisotropy.  Therefore, if reionization takes place later, the
anisotropy is significantly larger, even for the same optical
depth.  Furthermore, if reionization takes place later, the peak
of the anisotropy spectrum, which is fixed in comoving distance
scale by the peak of the power spectrum, subtends a larger
angle. 
In open models, the peak of the power spectrum moves to larger physical
scales, while the effects of geometery shift these physical scales back to
smaller angular scales; the result is that the spectrum peaks
at only slightly lower $\ell$ than in a flat universe.

The weakness of the dependence of the anisotropy amplitude on
$\tau_r$ indicates that a good fraction of the anisotropy is
produced at {\it later} times when the amplitude of the power
spectrum has grown.  The convergence of the predictions at small
$\ell$ for ionization histories with the same $x_e$ indicates
that, in particular, the anisotropy at larger scales comes
primarily from lower redshifts \cite{huwhite}, and therefore
that assuming that re-scattering occurs primarily at
the epoch of reionization\cite{efstathiou,dodelson} does not
provide a good approximation,
especially at larger scales.
Moreover, in open models the
evolution of the growth factor differs from a flat Universe by
larger amounts at later times; therefore the slope of $C_\ell$
at larger scales is greater than in a flat universe.

The growth of density perturbations induces an initially Gaussian
distribution of perturbations to become non-Gaussian.
Therefore, the distribution of anisotropies produced by the
Ostriker-Vishniac effect will be non-Gaussian even for Gaussian
initial conditions, and the higher $n$-pt temperature
correlation functions may be calculated in perturbation theory
\cite{workinprogress}.

The anisotropy is due primarily to density perturbations on
distance scales closer to those probed by galaxy surveys than to
those probed by {\sl COBE}.  Furthermore, the anisotropy is due
to a large extent to scattering at later times.  Therefore, when
the power spectra
are normalized to the cluster abundance, the spread in predicted
values for the rms anisotropy is smaller than it is if the
models are normalized to {\sl COBE}.   Since the anisotropy is a
second-order effect, the rms anisotropy is proportional to
$\sigma_8^2$.  In a flat Universe, models  normalized to
cluster abundances produce rms temperature anisotropies listed in Table
\ref{tab:predictions} of $0.8$--$3$ $\mu$K for a constant
ionization fraction of unity, whereas an ionization fraction of
0.2 yields rms anisotropies of $0.3$--$0.8$ $\mu$K.

Our numerical results can also be easily scaled for a different
value of the ionization fraction $x_e$ if the ionization epoch
is held fixed and if $\tau_r \ll 1$.  The rms temperature
anisotropy should be roughly proportional to $x_e$ if $z_r$ is
held fixed.  Also, if we neglect the effect of $\Omega_b$ and
$h$ on the power spectrum (in particular, $\Omega_b$ should have
only a relatively weak effect on the power spectrum as long as
$\Omega_b \ll \Omega_0$), then Eq.~(\ref{eq:tauequation}) shows
that the rms anisotropy should be proportional to the
combination $\Omega_b h x_e=(\Omega_b h^2) x_e/h$, itself proportional
to the column density of electrons to the last scattering
surface.  This also suggests that the Ostriker-Vishniac
anisotropy in models with a high baryon density
should be larger.  To illustrate (and to
illustrate the effects
of geometry), we also include in Table \ref{tab:predictions} and
in Fig. \ref{fig:Clsplot} predictions for an open model with
$\Omega_b \simeq \Omega_0/3$, as suggested possibly by a peak in
the observed power spectrum at 100~$h^{-1}$ Mpc
\cite{eisenstein}.  High-baryon-density models will be explored
more carefully in Ref. \cite{jaffeetal}.

\section{Approximate Vishniac Power Spectra for CDM Models}

Here we provide simple approximate analytic fits to the Vishniac
power spectrum $S(k)$ for standard-CDM power spectra.  These can
be used to quickly estimate the Ostriker-Vishniac anisotropy for
CDM models with an arbitrary ionization history.

With the standard-CDM power spectrum given in the form of
Eq.~(\ref{eq:powerspectrum}), $S(k)$ may be written,
\begin{equation}
     S(k) = {\pi^4 \delta_H^4 \over 16}\, (k/2)^{2n}\, k  \,
     I_n(k_p\,{\rm Mpc}/h \Gamma),
\label{eq:Sequation}
\end{equation}
where
\begin{equation}
     I_n(q) = \int_0^\infty dy \int_{-1}^1 dx
     (y\sqrt{1+y^2-2xy})^n T^2(yq) T^2(q\sqrt{1+y^2-2xy}) {
     (1-x^2) (1-2xy)^2 \over (1+y^2-2xy)^2}.
\label{eq:Ineqn}
\end{equation}
For the standard-CDM transfer function,
Eq.~(\ref{eq:transferfunction}), the functions $I_n(q)$ may be
approximated, for $0.5\lesssim n \lesssim 1$, by an asymmetric
Gaussian:
\begin{equation}
     I_n(q) \simeq {h(n) \over 4(q/2)^{2n+3}} \exp \left\{ -{1
     \over 2} \left[ {
     \log_{10}[q]-\log_{10}[q_0(n)] \over \sigma(n) } \right]^2 \right\},
\label{eq:Iapprox}
\end{equation}
with $\log_{10}[h(n)]=-4.716-0.636(n-1)+0.832(n-1)^2$,
$\log_{10}[q_0(n)] = 0.477+1.36(n-1)+0.975(n-1)^2$, and
$\sigma=0.6+0.33(n-1)+0.34(n-1)^2$ for $q<q_0$, and
$\sigma=0.87+1.105(n-1)+1.19(n-1)^2$ for $q>q_0$.

\begin{figure}[htbp]
\centerline{\psfig{file=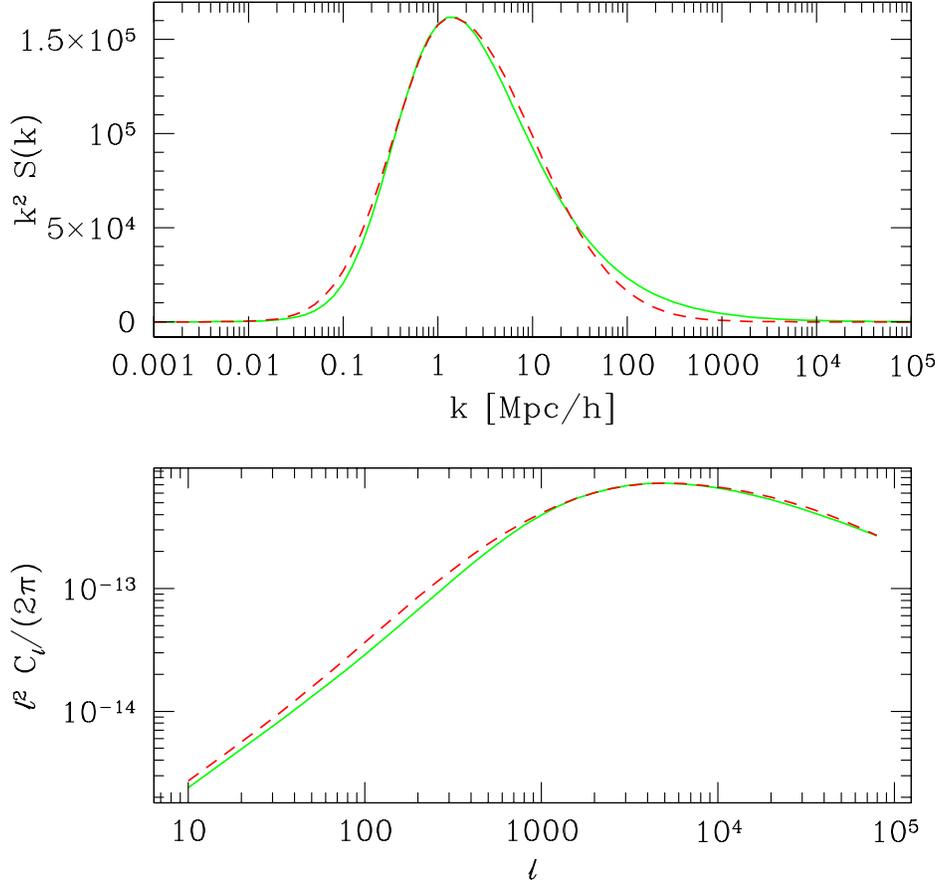,width=5in}}
\bigskip
\caption{
  Semi-analytic fits to $S(k)$, the Vishniac power spectrum. Top: full
  calculation (solid) and the analytic approximation (dashed) for the
  standard-CDM $S(k)$.  Bottom: $\ell(\ell+1)C_\ell/(2\pi)$ calculated
  with the full $S(k)$ (solid) and the approximation (dashed) for standard
  CDM with an ionization history given by $z_r=19$, $x_e=1.0$,
  $\tau_r=0.1$.  }
\label{fig:approxClsplot}
\end{figure}

Fig.~\ref{fig:approxClsplot} shows the multipole moments obtained
{}from numerical calculation of $S(k)$ and the analytic
approximation given here for a variety of models.  Although this
analytic approximation does not accurately reproduce the
asymptotic large- and small-$k$ behavior of $S(k)$, it accurately approximates
the peak of $k^2 S(k)$.  As indicated by
Eq. (\ref{eq:gaussianvariance}), this should provide reliable
estimates of the mean-square anisotropy.

\section{DISCUSSION}

Although precise limits depend on imprecisely determined
cosmological parameters, the existence of significant CMB
anisotropy at the degree scale and current limits to
arcminute-scale anisotropy and distortions to the CMB blackbody
spectrum seem to indicate that $\tau_r\lesssim 1$.  However,
some reionization is generically expected in CDM models.  Again,
the detailed ionization history will depend on the model and on
details of star formation that are still difficult to quantify
reliably.  Still, optical depths $\tau_r={\cal O}(0.1)$ are
observationally consistent and expected in CDM models.

Such a level of reionization will lead to secondary anisotropies at
arcminute scales like those shown in Section V. Furthermore, the bulk of
the reionization in CDM models generally occurs later, so the secondary
anisotropy will be larger for a given $\tau_r$ than if reionization
occurred earlier. The predicted level of anisotropy in virtually
all of the models we considered here is smaller than the current
limit of about 50 $\mu$K at arcminute angular separations
\cite{subrahmanyan}.  However, there has been much recent
progress in the design of interferometer experiments dedicated
to CMB measurements, and there should be much more in
forthcoming years.  To
estimate the signal that might be detectable, we
suppose that a detector with a sensitivity $s$ (in $\mu$K~$\sqrt{\rm
  sec}$) maps a fraction $f_{\rm sky}$ of the sky with an angular resolution
$\theta_{\rm fwhm}\lesssim 1'$ sufficient to detect these anisotropies for
a time $t_{\rm mon}$ months.  Such a detector should be sensitive to an
rms temperature anisotropy as small as
\begin{equation}
     (\Delta T)_{\rm rms}^{\rm min} \simeq 0.5\, \mu{\rm
     K}\,
     { (s/1000\, \mu{\rm K}\,\sqrt{\rm sec}) \, f_{\rm
     sky}^{1/4} \over (\theta_{\rm fwhm} /0.1^\circ)^{1/2}\, t_{\rm
     mon}^{1/2} }.
\label{eq:sensitivity}
\end{equation}
Although the interferometry experiments that will likely reach these
angular scales are not really parameterized by a sensitivity $s$, the
effective value of $s$ for these experiments is already better
than 1000 $\mu$K~$\sqrt{\rm sec}$.  Furthermore, with a full
power-spectrum analysis of the data, the sensitivity could be improved
perhaps by an order of magnitude over this simple estimate.  Comparing
Eq.~(\ref{eq:sensitivity}) for a reasonable $t_{\rm mon}$ with the
predictions in Section V, it seems quite plausible---even though the
experiments may be challenging---that the Ostriker-Vishniac anisotropy
expected in many leading CDM models will be detectable. The results of
Ref.\cite{interferometers} imply that the anisotropy generated by some of
the more optimistic models considered here may already be
detectable by upcoming experiments such as CBI, especially if
binned over a wide range of $\ell$.

Observation of the Ostriker-Vishniac effect will be important for
understanding the ionization history of the Universe and for determining
cosmological parameters with the CMB.  The linear-theory anisotropies
depend primarily on the optical depth $\tau_r$ and only secondarily on the
details of the ionization history (e.g., the values of $z_r$ and $x_e$).
Therefore, degree-scale anisotropies may determine $\tau_r$, but they will
not strongly constrain the epoch of reionization.  When combined with
sensitive measurements of the degree-scale anisotropy, measurement of the
Ostriker-Vishniac effect will therefore provide the epoch of reionization.

The shape of the CMB power spectrum for these anisotropies is relatively
featureless.  Therefore, even detection of the anisotropy, without a
precise determination of the shape of the power spectrum, will provide the
most essential information on the ionization history.  Of course, more
precise constraints to the ionization history can be obtained if the power
spectrum can be mapped and, some shape information along with
multi-frequency observations will be necessary to distinguish the
Ostriker-Vishniac anisotropy from foreground contamination such as that
induced by extragalactic point sources.

Determination of the geometry of the Universe relies primarily on the
first acoustic peak in the primary CMB spectrum
\cite{kamspergelsug,jkksone}.  However, precise and reliable
determination of other cosmological parameters, such as the baryon
density, Hubble constant, cosmological constant, and spectral index,
depends on the structure of the smaller-scale acoustic peaks.  The
details of the ionization history (e.g., the epoch of reionization,
which we have explored here) will, in fact, have some effect on these
higher peaks.  This will in turn affect the reconstruction of some
cosmological parameters, although this uncertainty has not yet been
taken into account in the parameter-estimation analyses of
Refs.\cite{jkks}.  Therefore, the information on the epoch of
reionization provided by mapping the secondary anisotropies will be
important for precise determination of these other cosmological
parameters with MAP and Planck.

It seems unlikely that the current FIRAS constraints to Compton-$y$
distortions of the CMB blackbody spectrum will be improved
significantly.  Therefore, the Ostriker-Vishniac effect seems to be the
most promising probe of the ionization history.  In addition to its
significance for cosmological-parameter determination, the
Ostriker-Vishniac effect can provide a window to the epoch of the
earliest collapsed objects that is inaccessible with any other
observations.

\acknowledgments

We thank M. White and W. Hu for useful comments.  This work was
supported at Columbia by D.O.E. contract
DEFG02-92-ER 40699, NASA NAG5-3091, and the Alfred P. Sloan
Foundation and at Berkeley by NAG5-6552. MK would like to acknowledge the
hospitality of the Center for Particle Astrophysics where part of this
work was completed.

\section*{APPENDIX: Projected Power Spectrum for a Divergence-Free Vector
Field}

If an isotropic random field $\vecq(\vecw)$ is
decomposed into Fourier components,
\begin{equation}
    \tilde\vecq(\veck) = \int \,d^3 \vecw\,\exp(-i \veck
     \cdot \vecw)\,      \vecq(\vecw),
\label{eq:vectorfourier}
\end{equation}
then the Fourier components may be written as the sum of
divergence-free and curl-free parts parts: $\tilde\vecq(\veck)=
\tilde\vecq_\perp(\veck) + \tilde\vecq_\parallel(\veck)$ with
$\veck \cdot \tilde \vecq_\perp(\veck) =0$ and $\veck \times
\tilde \vecq_\parallel(\veck)=0$; in particular,
$\vecq_\parallel(\veck)=\hatk q_\parallel(\veck)$.

Suppose we observe a projection
\begin{equation}
     p(\vectheta) = - \int_0^{\eta_0} \, dw \,
     g(w)\, \hattheta \cdot \vecq(w\hattheta,w;w),
\label{eq:intermediateonetwo}
\end{equation}
of $\hattheta \cdot \vecq(w\hattheta,w)$, the component of
$\vecq(\vecw)$ along the line of sight, with a visibility function
$g(w)$, and we define the Fourier components,
\begin{equation}
     \tilde p(\veckappa) = \int\, d^2 \vectheta\, p(\vectheta) \,
     e^{- i \veckappa \cdot \vectheta}.
\end{equation}
The contribution of the parallel components $\tilde
\vecq(\veck)$ to $p(\vectheta)$ will be small because crests and
troughs of each Fourier component projected along the line of
sight will tend to cancel.  The purpose of this Appendix is to
show that if the perpendicular components satisfy $\VEV{\tilde
\vecq_\perp(\veck) \cdot \tilde \vecq_\perp^*(\veck') } =
(2\pi)^3 \delta(\veck-\veck') P_\perp(k)$, then the Fourier
components of the projection satisfy
\begin{equation}
     \VEV{\tilde p(\veckappa) \tilde p^*(\veckappa')} = (2\pi)^2
     \delta(\veckappa - \veckappa') P_p(\kappa).
\label{eq:VEVeqn}
\end{equation}
with
\begin{equation}
     P_p(\kappa) = {1\over 2}\int_0^{\eta_0}\, dw\, {g^2(w)
     \over w^2} P_\perp(\kappa/w,w).
\label{eq:kaiserlimber}
\end{equation}
To do so, we follow the steps in the Appendix of
Ref. \cite{kaiser} that lead to the Fourier-space analog of
Limber's equation.
Rather than reconstruct the entire argument, we refer the
reader to Kaiser's paper \cite{kaiser} for more details.

The components of an isotropic divergence-free field satisfy
\cite{MoninYaglom},
\begin{equation}
     \VEV{\tilde q_{\perp,i}(\veck) \tilde
     q_{\perp,j}^*(\veck')} = {1\over 2} (\delta_{ij} - \hatk_i
     \hatk_j) \VEV{ \tilde \vecq_\perp(\veck) \cdot \tilde
     \vecq_\perp^*(\veck')}.
\label{eq:moninequation}
\end{equation}
If the contribution of
a narrow shell of width $\Delta w$ centered at $w_0$ is
\begin{equation}
     \Delta p(\vectheta) = g(w_0) \int_{w_0-\Delta
     w/2}^{w_0+\Delta w/2}\, dw\, \hattheta \cdot \vecq_\perp(w_0
     \theta_1, w_0 \theta_2,w; w),
\end{equation}
then with Eq.~(\ref{eq:moninequation}), we can show that its
Fourier coefficients $\widetilde{ \Delta p}(\veckappa)$ satisfy
\begin{equation}
     \VEV{\widetilde{\Delta p}(\veckappa)\, \widetilde{ \Delta
     p}^*(\veckappa')} = (2\pi)^2 \delta(\veckappa - \veckappa')\,
     \Delta P_p(\kappa).
\end{equation}
with
\begin{equation}
     \Delta P_p(\kappa) = {1\over 2} {g^2(w_0) \Delta w \over w_0^2}
\int {dk_3 \over 2\pi} [1 - (\hattheta
     \cdot \hatk )^2] [j_0(k_3 \Delta w/2)]^2 P_\perp \left(
     \sqrt{ {\kappa^2 \over w_0^2} + k_3^2} \right),
\label{eq:kaiserlimbertwo}
\end{equation}
and $\veck = (\kappa_1/w_0,\kappa_2/w_0,k_3)$.  As argued by
Kaiser, only modes with $k_3 \ll \kappa/w_0$ (those very nearly
perpendicular to the line of sight) will contribute appreciably
to the integral, so $(\hattheta \cdot \hatk)^2 \ll 1$.  We
then add the contributions $\widetilde{\Delta p}(\kappa)$ along the line of
sight to obtain Eq.~(\ref{eq:kaiserlimber}).


In an open or closed Universe one replaces
\begin{equation}
      w \quad \longrightarrow \quad { {\cal S}(a_0 H_0
      w\sqrt{|1-\Omega_0-\Omega_\Lambda|})
        \over a_0 H_0 \sqrt{|1-\Omega_0 -\Omega_\Lambda|}}
\label{eq:replacement}
\end{equation}
in the first argument of $P(\kappa/w,w)$ in
Eq.~(\ref{eq:kaiserlimber}), and in the two factors of $w$ that
appear in the denominator, where ${\cal S}(x)=\sinh x$ in an
open Universe and ${\cal S}(x)=\sin x$ in a closed
Universe \cite{kaiser96}.\footnote{We thank Wayne Hu for
pointing out an error in this scaling in a previous draft.}

Note that this derivation also explains why the contribution from
$\vecq_\parallel$ is suppressed. This term satisfies \cite{MoninYaglom}
\begin{equation}
  \VEV{\tilde q_{\parallel,i}(\veck) \tilde q_{\parallel,j}^*(\veck')}
  = {\hat k}_i {\hat k}_j \VEV
  {\tilde q_\parallel(\veck)\tilde q_\parallel^*(\veck')}
  = (2\pi)^3\delta(\veck-\veck'){\hat k}_i {\hat k}_j P_\parallel(k).
  \label{eq:secondmoninequation}
\end{equation}
This would therefore have a contribution to the projected
power spectrum of $\hattheta\cdot\vecq$ proportional to
$(\hattheta\cdot\hatk)^2\ll1$.

The approximation used here is valid for small angular separations,
$\kappa \gg 1$.  Since the Ostriker-Vishniac effect produces
anisotropies at multipole moments $\ell\gtrsim 1000$, this flat-sky
approximation should lead to errors no greater than $O(0.1\%)$. At large
angular scales, we can also expect the anisotropy from the parallel term
to be significant\cite{huwhite}.  Also, the visibility function $g(w)$ should
be smoothly varying over wavelengths at which there is significant
power.  Again, the visibility function for re-scattered CMB photons will
be very broad, so this approximation should be fine for our calculation.

\end{document}